# Comment on "Possible Precise Measurement of Delbrück Scattering Using Polarized Photon Beams"

Sylvian Kahane
P. O. Box 1630, Omer, Israel, 8496500
Retired from Phys. Dept., NRCN, Israel

In their Letter[1] to PRL (with the above title), Koga and Hayakawa presented an over optimistic scenario in which, due to the destructive interference between tree components (Rayleigh, Thomson (nuclear) and GDR (Giant Dipole Resonance)) of the elastic photon scattering from the field of a nucleus, an extremely accurate measurement of the fourth component – the Delbruck scattering, will be possible in the near future, using polarized photons from the laser Compton scattered facilities. They claim that the optimal energy for such a measurement will be close and a bit above 1 MeV. Delbruck scattering is an extremely interesting nonlinear phenomenon, belonging to the class of photon-photon scattering, and of the minimum 4th order in QED.

My principal objection to the calculations of Koga and Hayakawa is in the use of photons of 1 MeV energy and the estimate of GDR contribution, at this energy, by employing the Lorentzian parametrization of GDR. A mistaken evaluation of the GDR contribution will destroy the claim of almost total destructive interference between R, T and GDR, and by extension, of an almost pure measurement possibility of Delbruck scattering.

It is well known that the parametrization to the GDR is obtained by fitting one or two Lorentzian-s to the measurements of the ($\gamma$,n) cross sections. The experimental data begin to appear at energies substantial larger than 1 MeV, due the binding energy of the neutron. For example, Fig. 1 presents the experimental data for $^{88}$Sr[2]:

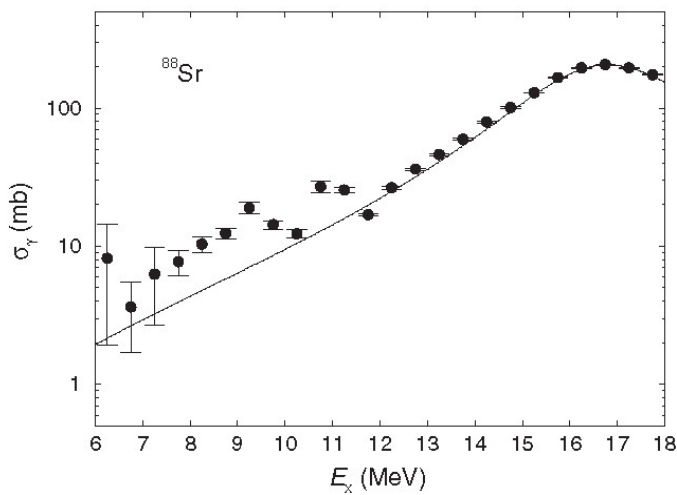

Fig. 1. GDR of $^{88}$Sr from ref. 2.

The solid line is the fitted Lorentzian, by which the prediction, for the cross section at 6 MeV is 2 mb, while the measured value is close to 8 mb. It is expected that an extrapolation to 1 MeV will give very wrong results.

But the problem is deeper. As the name suggests, the GDR implies nuclear excitations of the dipole type. Lately, a large number of experiments, mainly at the university of Oslo, probed the $\gamma$-strength function in the low energy region of a couple of MeV, using a number of different reactions and methods (one especially known as the "Oslo Method"), in particular not using the ($\gamma$,n) reaction. A major finding of these works is that contributions are made, at these energies, also by other multipolarities and by a peculiar upbend of the strength function (and hence by cross sections) at and below 1 MeV, see Fig. 2:

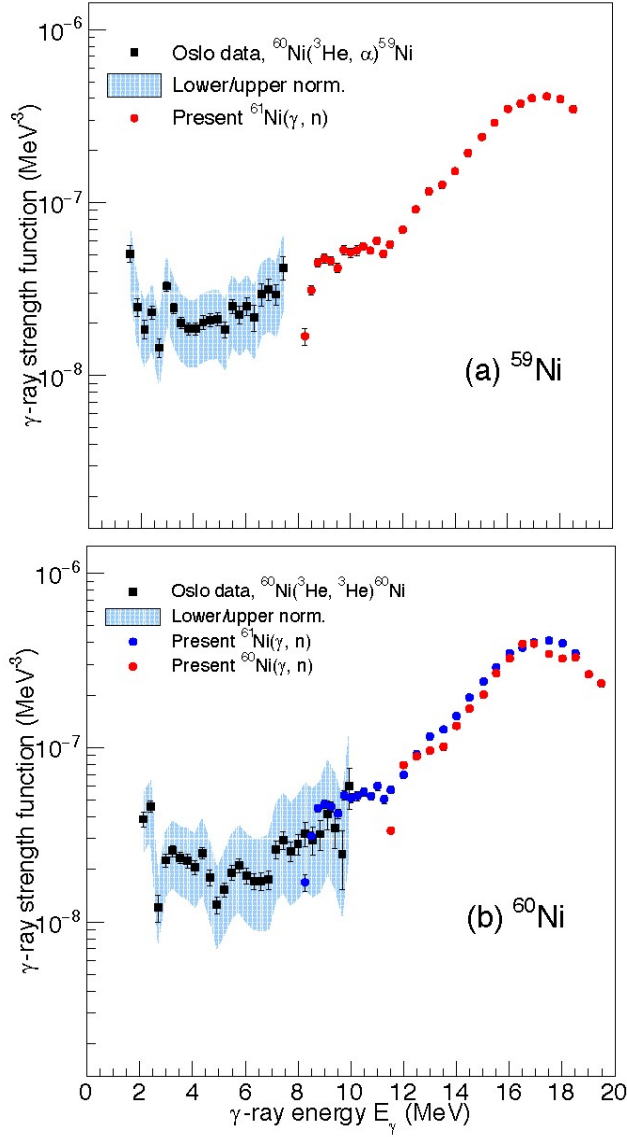

FIG. 8. (Color online) Combining the γSF above and below $S_n$ for $^{60}$Ni in (a). In (b) both γSFs from $^{61}$Ni and $^{60}$Ni above $S_n$ are combined with the γSF of $^{59}$Ni for the region below $S_n$.

*Fig. 2. Data for γSF of $^{59\text{-}60}$Ni, taken from Ref. 3.*

Only the red points were measured in (γ,n) reactions and contribute to the Lorentzian parametrization. It is clear that the lower points will be missed by such a parametrization.

I believe that more viable option is to use a higher energy and a forward scattering angle. Delbruck scattering is described by four amplitudes: real and imaginary, with photon polarization parallel or perpendicular to the scattering plane (or alternatively with or without spin-flip). The imaginary amplitudes are related with the absorption process of pair production and not of a particular interest in this context. The prize stays in the real amplitudes related with the exotic process of vacuum polarization. In a measurement with polarized γ-rays only two amplitudes will contribute. The following graph (Fig. 3), based on the calculations of Ref. 4 up to 28 MeV and of Ref. 5 up to 70 MeV, presents the dominant real parallel Delbruck amplitudes, as a function of energy, at a scattering angle of 1°. It will be beneficial to perform a measurement using 15 – 16 MeV γ rays.

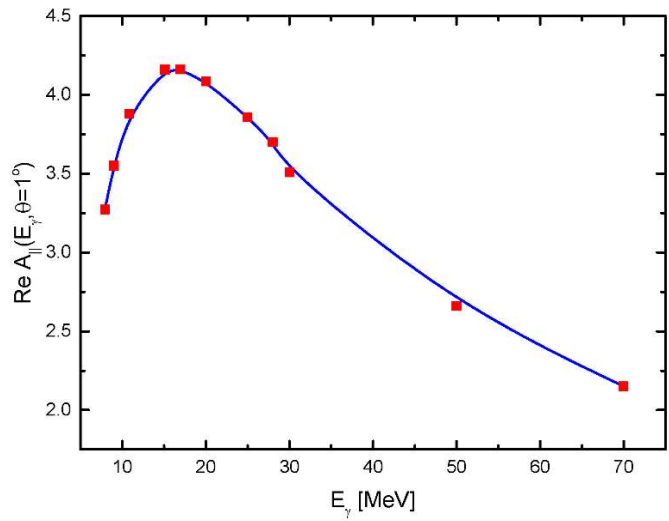

*Fig. 3. The real part parallel Delbruck amplitudes.*

At these energies the real parallel Delbruck amplitude is higher by a factor of more than 400 compared with the one at 1 MeV (compare with Fig. 1 of Ref. 1). At such an energy there will be a sizable contribution also from the imaginary amplitude. Both will enhance the cross sections at high values permitting shortening the measurement times from more than a month (as proposed by Koga and Hayakawa) to a couple of days or even less, making it much easier to obtain beam time. At the existing NewSubaru facility such $\gamma$ energies appear feasible (see Fig. 2 in Ref. 15 of the Letter). At the UVSOR (Ref. 16 of the Letter) the maximum $\gamma$ energy is 6.5 MeV. Even at this lower energy, extrapolating the above graph, the Delbruck amplitude will be a factor of about 200 higher compared with 1 MeV. The future ELI-NP-GBS facility (Ref. 17) will cover the range 1-20 MeV of $\gamma$ energies with a resolution (bandwidth in their parlance) of less than 0.5%, i.e., 75 keV at 15 MeV. This resolution is quite large compared with the existing measurements which employed neutron capture $\gamma$ rays with widths in the eV range and remains to be seen what will be the influence.

Using a forward scattering angle has additional pros and contras. Both R and D processes are very forward peaked when the energy increases. At the proposed energies, around 15 MeV, this feature makes the contributions of T and GDR (which both have a dipole angular dependence of $1+\cos^2(\theta)$) completely negligible. This was already tested at $\theta = 1.5°$ forward angle scattering in the work reported in Ref. 6. Besides, the proposed energies are close to the peak of the GDR where the Lorentzian shape is a good descriptor, hence no surprises connected with the tail. Moreover, small scattering angles imply small momentum transfers, and, according to Ref. 7, eliminating the contributions of the Coulomb corrections to the Delbruck scattering, for which no calculations are available. Still the R+D interference has to be accounted for, while R amplitudes calculated in the S matrix formalism are not available. What is left is the Modified Form Factor (MFF) approximation, in a single electronic configuration[8] or in a configuration interaction (CI) picture[9]. At small momentum transfers this approximation is expected to be pretty good. It was used extensively in a number of previous investigations.